# Solving viscoelastic problems in a step-inverse Laplace transform approach supplanted with ARX models: a way to upgrade Finite Element or spectral codes.

Stéphane André[1,2], Camille Noûs[2]

(1) Université de Lorraine, CNRS, LEMTA, F-54000 Nancy, France
(2) Cogitamus Laboratory, F-75000, France

**ABSTRACT:**

Finite Element codes used for solving the mechanical equilibrium equations in transient problems associated to (time-dependent) viscoelastic media generally relies on time-discretized versions of the selected constitutive law. Recent concerns about the use of non-integer differential equations to describe viscoelasticity or well-founded ideas based upon the use of a behavior's law directly derived from Dynamic Mechanical Analysis (DMA) experiments in frequency domain, could make the Laplace domain approach particularly attractive if embedded in a time discretized scheme. Based upon the inversion of Laplace transforms, this paper shows that this aim is not only possible but also gives rise to a simple algorithm having good performances in terms of computation times and precision. Such an approach, which fully relies on the Laplace-defined Behavioral Transfer Function (LTBF) can be promoted if it uses ARX parametric models perfectly substitutable to the real LTBF. They avoid the hitherto prohibitive pitfall of having to store all past data in the computer's memory while maintaining an equal computation precision.

# 1 Introduction

To be firmly characterized, the ViscoElastic (VE) behavior of solid materials requires a mathematical model associated to this rheology, as well as a metrological approach to correctly identify the involved physical parameters. For a Representative Elementary Volume (REV) of viscoelastic media, the model relates in a biunivocal correspondence both the stress and the strain applying onto this REV as a function of time. In the case of linear viscoelasticity, these variables are linked through a simple convolution product in time domain which kernel describes the VE behavior (relaxation or retardation depending on whether the strain or the stress are considered for the excitation). Through the convolution theorem, using Laplace or Laplace-Carson transforms of the behavior's model -one should call it the "transfer function" of the material- is a standard approach to compute the response (or output) of the material volume to a given time-dependent solicitation (or input) (Tschoegl, 1989). This is particularly true since the development of numerical algorithms that perform inverse Laplace transform precisely (Davies & Martin, 1979; de Hoog et al., 1982; Dingfelder & Weideman, 2014; Iseger, 2006; Stehfest, 1970; Talbot, 1979). The transfer function formally corresponds to the output obtained for a Dirac delta distribution (pulse) considered as input. Inherent to experiments in solid mechanics, the case of the impulse response in quasi-static[1] conditions cannot be obtained as in other scientific fields (the flash method for instance to measure the thermal diffusivity of materials, (Degiovanni & Jannot, 2018)). Therefore, creep, relaxation (Dooling et al., 1997; Emri & Tschoegl, 1993) or even indentation tests (Uluutku et al., 2022) are generally preferred to access a retardance or relaxance (Tschoegl, 1961), corresponding respectively to an admittance or to an impedance in the electrical analogy.

One interest of the Laplace formulation of VE behaviors is for the harmonic excitation which corresponds to the Dynamic Mechanical Analysis (DMA) experimental technique. In that case, the harmonic modelling corresponds to a reduction of the one-sided Laplace transformed problem where the Laplace variable ($p$ notation used in this article) is restricted to the imaginary axis $p = j\omega$, where $\omega$ is the harmonic pulsation of the input excitation. But although many DMA devices are produced by a few worldwide manufacturers, they can still be hardly used for metrological precise characterizations. Depending on the conducted loading test (bending, torsion, traction, shearing tests), it is still a research concern to obtain a perfect match of DMA-measured mechanical parameters and those reached with quasi-static experiments for other type of excitations (ideally with multisequence loading). The generally small specimen sizes, very weak considered forces and displacements in linear VE regime, add to this difficulty. These experiments are however widely used for qualitative purposes, when discrimination between different materials is looked for, on damping properties typically. DMA is considered to be a very precise tool to investigate phase transitions (Menard & Menard, 2020) for example but, in this case, experiments are rather conducted at a given frequency for a sweep in temperature (measurement of a glass transition typically).

---

[1] If inertial effects are taken into account, VE materials can be characterized through complex moduli measured with the vibrational response of beam-like structures where the excitation can be produced in forms of an impulsion (hammer blow).

Anyway it would be of great benefit, especially for viscoelastic heterogeneous materials (Agbossou et al., 1993; Shaterzadeh et al., 1998), to make the technique enters in a fruitful interaction with simulation tools (André et al., 2021; Fuentes et al., 2017; Gallican & Brenner, 2019; Trumel et al., 2019). The perspective that could be draw, is the following. Assuming that DMA could characterize more efficiently the behavior's law in terms of transfer function, and that this transfer function could be implemented directly in a FEM mechanical code through the behavior's law module, then simulations at the scale of a structure could be directly performed. The problem to circumvent is that numerical codes will necessarily proceeds with a time-step algorithm (an integration scheme in time) which appears incompatible with the use of a Laplace transform defined behavior's law. The present paper shows how this can be achieved with a different philosophy than that implemented in previous works on heterogeneous materials (Lévesque et al., 2007; Rekik & Brenner, 2011) or on homogeneous materials by using the classical collocation method (Schapery, 1962) to identify the relaxation kernel from its expression in Laplace domain.

We shall first (Section 2) recall the equations and vocabulary associated to the VE problem and define the central behavior's law considered for this study, which can be reduced to either the simple analogical Standard Linear Solid (SLS) behavior nor asymptotically, to a non-integer or fractional (NIF) order operator. In Section 3, the incremental approach is built by making use of a Laplace-defined behavior's law and of its response to a unitary ramp excitation. This solution is shown precise but involves all the history of the mechanical variables which makes it inefficient to use with meshed structures. ARX models' structure is shown here to be very powerful to avoid this drawback. Section 4 illustrates how it works and the achieved performances with the three types of VE behavior considered in this study.

## 2 Mathematical viscoelastic problem

### 2.1 Generic equations

The mathematical problem characterizing the viscoelastic behavior is of convolution nature through the well-known Boltzmann superposition principle (Christensen, 2012). The stress for example is given by the convolution product

$$\sigma(t) = \int_0^t H(t-\tau)\varepsilon(\tau)d\tau = \int_0^t G(t-\tau)\dot{\varepsilon}(\tau)d\tau \qquad (1)$$

where $H(t)$ figures the relaxance kernel (Tschoegl, 1989) and $\varepsilon(t)$ the test function i.e. the excitation (strain loading path) imposed on the system. Alternatively, in the second equality, $G(t)$ figures the relaxation function or relaxation modulus with $G(t) = \int_0^t H(\tau)d\tau$ and $H(t) = \frac{dG}{dt} + G(0)\delta(t)$ as a result of $G$ being discontinuous at $t = 0$ but with a definite limit. The lower integral bound initiates at $t = 0$ as for any causal problem. This expression suggests that the full-time history dependence of variable $\varepsilon$ (since the experiment starts) contributes to the actual value of the stress. The VE problem considered here in 1-D is referred to as a SISO model (Single Input-Single Output) with the vocabulary of control theory and referring to the role it plays in the historical development of the parametric models used below (Gevers, 2006).

To this mathematical formulation can correspond an alternative form in terms of Ordinary Differential Equation (Eq.2) through a linearity property. One can have the following general equation

$$\sum_{i=0}^n \alpha_i \sigma^{(i)}(t) = \sum_{j=0}^m \beta_j \varepsilon^{(j)}(t) \qquad (2)$$

which produces a given model structure for the relaxance kernel $H(t)$, readily obtained through Laplace transforming of Equation (2). Laplace-Carson transform is in general invoked in the mechanical community but is just a commodity practice to access directly the relaxation modulus (or alternatively creep function, considering a permutation of the independent variables $\sigma$ and $\varepsilon$ in Equation (1)) when a Heaviside function is considered for $\varepsilon(t)$: the measured stress is then the image of the time-dependent relaxation modulus. But it is not necessary for the reasoning and even blurs the full generality achievable with the transfer function concept. In Laplace domain, Equation (2) becomes

$$\mathcal{L}\{H(t)\} = \bar{H}(p) = \frac{\bar{\sigma}(p)}{\bar{\varepsilon}(p)} = \frac{\sum_{i=0}^n \alpha_i p^i}{\sum_{j=0}^m \beta_j p^j} \qquad (3)$$

$\bar{H}(p)$ can also be defined directly through a specific mathematical function and one can think here to the empirical Davidson-Cole relaxation model or to the more recently promoted models with non-integer exponents leading to fractional differential operators substituted to integer ones in Equation (2) (Bagley & Torvik, 1986; Schiessel et al., 1995). This aspect is mentioned here to draw attention of the reader to the large application that can be made from the results presented here. When $\bar{\varepsilon}(p) = 1$, $\bar{H}(p)$ is named the transfer function of the system. Using a metaphorical expression from biology, the transfer function characterizes the "full DNA" of the system (material) in terms of frequency information. It corresponds to the response of the system to the impulse (Dirac Delta) function $\bar{H}(p) = \overline{\sigma^\delta}(p)$ and is obviously never realized experimentally in mechanics due to an impossibility related to inertial effects (but is a quasi-systematic approach for thermal systems excited with laser pulses for example, as exemplified in (Corbin & Turriff, 2012)). When such transfer function is known from the engineer, then it allows computing the response of the system to any input excitation. If $\varepsilon(\tau) = \delta(t)$, the multiplicative identity of a convolution product,

$$\sigma^\delta(t) = \int_0^t H(t-\tau)\delta(\tau)d\tau = H(t) \qquad (4)$$

is the realization of the transfer function and then, in any other known specific strain **l**oading **p**ath $\varepsilon^{\ell p}(t)$, we have $\sigma(t) = (\sigma^\delta * \varepsilon^{\ell p})(t) = \int_0^t \sigma^\delta(t-\tau)\varepsilon^{\ell p}(\tau)d\tau$ fully determined by the impulse response $\sigma^\delta(t)$.

## 2.2 Description of the considered VE behavior's laws and loading sequences

For the computations presented later, we will consider as generic viscoelastic transfer function, the one furnished by the Dynamic of Linear Relaxations (DLR) viscoelastic model (Cunat, 2001). This model originates from a TIP framework (Thermodynamics of Irreversible Processes) combined to a modal approach of the dissipative processes. To say it shortly, this model ends in a finite set of ODEs identical to Equation (2) but where coefficients $\alpha_i, \beta_j$ are in auto-recursive geometric progression, or equivalently, when the Laplace counterpart of Equation (3) exhibits a recursive Pole and Zero Distribution. This model has been shown efficient to describe and measure associated material constants of HDPE, a thermoplastic, semi-crystalline polymer (Blaise et al., 2016).

This summarizes into

DLR constitutive equation

$$\dot{\sigma}(t) = \sum_{j=1}^{N} \dot{\sigma}_j = \sum_{j=1}^{N} \left( p_j \cdot E^u \cdot \dot{\varepsilon} - \frac{\sigma_j - p_j \cdot \sigma^r}{\tau_j} \right) \quad (5.1)$$

With
$E^u$ : **u**nrelaxed (Young or glassy) Modulus,

- $(p_j, \tau_j)$: the spectrum of relaxation times defined through $\tau_j = \tau_{max} \cdot 10^{-\left(\frac{N-j}{N-1}\right)d}$ ($\tau_{max}$: material parameter corresponding to the maximum relaxation time, d and N respectively the number of decades and of relaxation modes of the spectrum) and the weights $p_j^0 = \frac{\sqrt{\tau_j}}{\sum_{j=1}^{N} \sqrt{\tau_j}}$ ensuring $\sum_{j=1}^{N} p_j^0 = 1$ with the recursion $p_{j+1}/p_j = \alpha$ ; $\tau_{j+1}/\tau_j = \beta$.
- $\sigma^r(t)$ : The **r**elaxed state defined simply as $\sigma^r(t) = E^r \varepsilon(t)$ where $E^r$ is the **r**elaxed modulus (rubber modulus for polymers)

DLR constitutive equation in Laplace Domain (LBTF)

$$\bar{H}(p) = \frac{\bar{\sigma}(p)}{\bar{\varepsilon}(p)} = \sum_{j=1}^{N} p_j \left( \frac{\tau_j E^u p + E^r}{1 + \tau_j p} \right) \quad (5.2)$$

DLR impulse response in time (Transfer function)

$$\sigma^\delta(t) = \mathcal{L}^{-1} \bar{H}(p)(t) = \sum_{j=1}^{N} p_j \cdot E^u \delta(t) + \sum_{j=1}^{N} p_j (E^r - E^u) \frac{e^{-t/\tau_j}}{\tau_j} \quad (5.3)$$

Two other transfer functions will be considered later that originates from this generic one referred to next as behavior (B). All of them will participate to a validation of the incremental approach described next and the understanding its underlying subtilities.
- One is the Standard Linear Solid (SLS) or 3-parameter Voigt model and corresponds to the reduction of the DLR model to one single relaxation time. The transfer function and other formulations of the SLS behavior's law are directly obtained from equations (5.1-3) by discarding the summation operator.
- The other one is obtained when considering asymptotically the case of an infinite number of modes ($N \to \infty$) which we have shown earlier to be related to a specific fractional model (André et al., 2003) i.e. a model based on non-integer derivative operators. It is obtained directly in Laplace domain without any analytical expression of its original and will be described later in section 4.3. This case is considered here in view of the abundant literature available on these NIF (Non-Integer or Fractional) models since the 90's as they allow for the description of a plethoric set of VE behaviors. Also, their natural definition in Laplace domain will legitimate even more deeply the tool raised by this study in view of an introduction into numerical codes. Because this fractional behavior stems from the DLR behavior and remains rather confidential, a second fractional law based on Rabotnov's suggestion of using a fraction-exponential operator, widely considered since 1948 (Koeller, 1984; Sevostianov et al., 2015), will also be quickly investigated in this section. Whatever "fractional" behavior considered, it will be referred next as a NIF model.

For the simulations presented later, we will also consider a set of different loading paths $\varepsilon^{\ell p}(t)$ presented in Table 1 along with their known Laplace domain expression. Case 0 corresponds to a smooth function in time. Case 1 corresponds to the crenel excitation and Cases 2 to 4 correspond to more complex sequences made of ramps. The associated graphs of these loading paths (inputs on the system) will be shown later along with the calculated responses (outputs).

| $\varepsilon^{\ell p}(t)$: | Time domain | Laplace domain |
|---|---|---|
| Case 0 | $\dfrac{2}{\sqrt{3}} e^{-t/2} \sin\left(\dfrac{\sqrt{3}}{2} t\right)$ | $\dfrac{1}{s^2 + s + 1}$ |
| Case 1<br>Crenel | $\begin{cases} \varepsilon_0 & 0 \leq t \leq t_c \\ 0 & t > t_c \end{cases}$<br><br>$t_c = 2s$ and $\varepsilon_0 = 1$ in the simulations | $\varepsilon_0 \left(\dfrac{1}{s} - \dfrac{1}{s} e^{-st_c}\right)$ |
| Case 2<br>2 successive ramps | $at + b(t - t_{r1})\mathcal{H}(t - t_{r1})$<br><br>$t_{r1}$: time at 1st slope change in ramp<br>$b$: slope change (means that the 2nd ramp will have a slope $a + b$) | $\dfrac{1}{s^2}(a + be^{-st_{r1}})$ |
| Case 3<br>3 successive ramps | $at + b(t - t_{r1})\mathcal{H}(t - t_{r1})$<br>$+ c(t - t_{r2})\mathcal{H}(t - t_{r2})$<br><br>$t_{r2}$: time at $2^{nd}$ slope change in ramp<br>c: new slope change | $\dfrac{1}{s^2}(a + be^{-st_{r1}} + ce^{-st_{r2}})$ |
| Case 4<br>3 successive ramps and stabilization | $at + b(t - t_{r1})\mathcal{H}(t - t_{r1})$<br>$+ c(t - t_{r2})\mathcal{H}(t - t_{r2})$<br>$- d(t - t_{stab})\mathcal{H}(t - t_{stab})$<br><br>$t_{stab}$: time at which a dwell is imposed to the strain<br>$d = (a + b + c)t_{stab}$ | $\dfrac{1}{s^2}(a + be^{-st_{r1}} + ce^{-st_{r2}} - de^{-st_{stab}}) -$ |

**Table 1:** Considered test cases for the loading strain path ($\mathcal{H}(t)$ : Heaviside function)

# 3 Towards an incremental time-step approach of the problem

## 3.1 The step-approach reached from the unit ramp response

FE codes in solid mechanics and more recent FFT-based spectral solvers are up-to-know based on a time-discretization approach for elasto-visco(-plastic) materials submitted to transient excitation. They do not support a Laplace definition of the transfer function. The equilibrium equations are solved at specific locations of the spatial discretized domain. For each increment in time step, a call to the behavior's law computation is required which is always specified in time discretized version, with different possible schemes (Sorvari & Hämäläinen, 2010). The objective of the present study is to allow the behavior law of the material be defined directly in the Laplace domain. Subsequently, this may offer a means of solving viscoelastic problems starting from experimental characterizations made by DMA to then predict the expected behaviors in all other conditions.

We first explain how to compute precisely and *incrementally* the temporal response of a viscoelastic material with its Laplace-defined Behavioral Transfer Function (LBTF). By 'incremental' we precisely mean running a loop over a time-discretized vector and calculating the output stress according to $\sigma(t + \Delta t) = \sigma(t) + \Delta\sigma^t$ where the challenge is to obtain an estimation of $\Delta\sigma^t$. For the different test-cases considered, the incremental computation referred to as **step-LBTF**, will be compared in the whole-time interval $[0; t_f]$ with

(i) A full inverse Laplace Transform approach, meaning that the computation will be based on the definition of the Laplace Transformed behavior's law (Eq. 5.2) but applied to the whole discretized vector of the time domain under consideration $[0; t_f]$. We will use the Laplace inversion based on De Hoog's algorithm (de Hoog et al., 1982). These computations will be labelled in the figures as **full-LBTF**. Note that if nothing proscribes the use of De Hoog's algorithm embedded in a time loop to calculate *iteratively* the output stress, although in largely increased CPU times but generally with greater precision, this is not our objective here because we are seeking for an iterative but *incremental* algorithm, in view of local calculations performed in FEM or Spectral solver codes.

(ii) The computation of the response through the convolution product over the full-time interval (Eq. 5.3) between the transfer function (impulse response $\sigma^\delta(t)$) - when known mathematically - and the input excitation (strain loading path $\varepsilon^{\ell p}(t)$). The result will be referred to as **Convolution** $(\sigma^\delta * \varepsilon^{\ell p})(t)$.

The incremental computation over time steps $\Delta t = [t_k; t_{k+1}]$ will be based on the unitary ramp response assumed to be known at the discretized times of the computation and the knowledge of the discretized slope variations $\Delta S_k = S_k - S_{k-1}$ at time $t_k$ (Figure 1) with $S_k = \left(\varepsilon^{\ell p}(t_{k+1}) - \varepsilon^{\ell p}(t_k)\right)/\Delta t$.

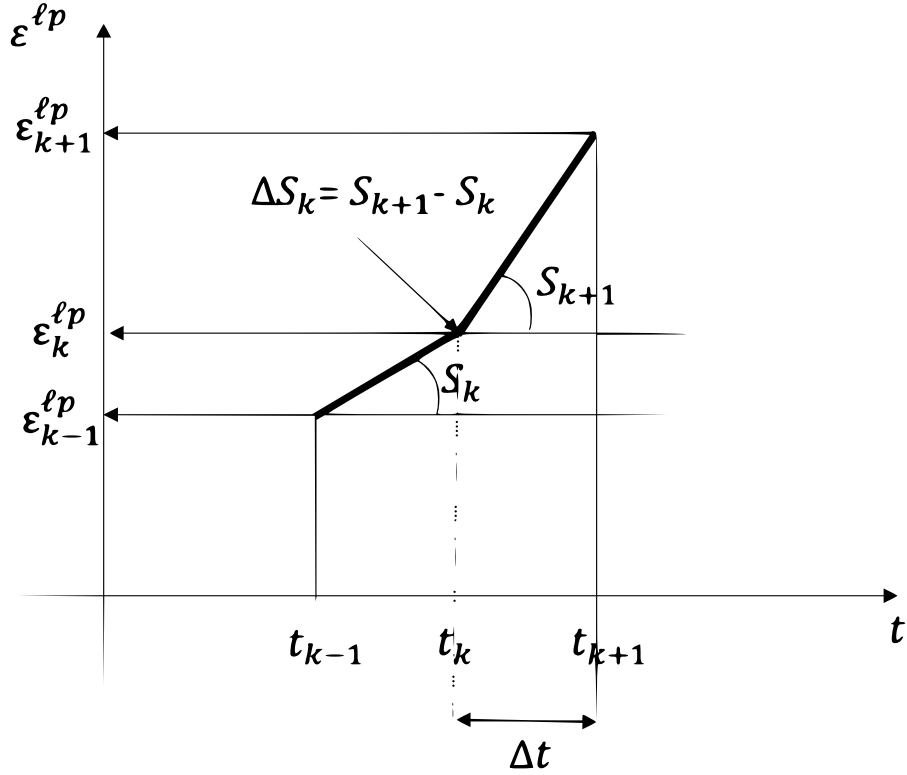

**Figure 1:** Discretization of the input $\varepsilon^{\ell p}$ in terms of slope variations.

As a result, the algorithm relies on a numerical approach built according to Duhamel's theorem (Özisik, 1993; Sneddon, 1951).

The first step is to compute the elementary time response to a Unitary Ramp (ramp of unit slope over $[0; \Delta t]$) which then remains *Stucked* at a Stationary value equal to $1 \times \Delta t$. This input is described as: $\varepsilon^{\ell p} \begin{cases} 0 & t < 0 \\ t & 0 \leq t \leq \Delta t \\ \Delta t & t > \Delta t \end{cases}$. We named it hereafter URSS-Response and noted it $\sigma_0^{\nearrow-}$ with subscript '0' referring to the time origin when this excitation takes place.

It can be computed once for all in a preliminary stage using for example the Laplace inverse given in the rhs of Equation (5.4) and stored as a vector of the discretized values $\sigma_0^{\nearrow-}(t_k)$ over all time steps.

$$\sigma_0^{\nearrow-}(t) = \mathcal{L}^{-1}\left\{\bar{H}(p)\frac{1}{p^2}(1 - e^{-p\Delta t})\right\} \qquad (5.4)$$

The second step consists in solving the problem $\sigma(t) = \mathcal{L}^{-1}\{\bar{H}(p)\varepsilon^{\ell p}(p)\}$ in terms of the incremental scheme $\sigma(t + \Delta t) = \sigma(t) + \Delta\sigma^t$ where $\Delta\sigma^t$ will be computed from the URSS-Response (see Appendix). Indeed, the use of the time translation invariance property allows to compute $\Delta\sigma^t$ as the "cumulative" dot product of the backward URSS-Response originating recursively from time $t$ with the vector of slope increments up to time $t$. The first unitary ramp of slope $\Delta S_1$ acts on the VE response over all subsequent time steps. The second unitary ramp of slope $\Delta S_2$ acts on the VE response delayed from one time-step, and so on. This corresponds to the algorithm given in the Appendix. Classically, the mid-point approach enhances the precision.

Figure 2 compares the results obtained with the 3 approaches: full-LBTF, convolution, Step-LBFT in the case where the relaxation spectrum is limited to one mode (Case 1, SLS rheological element). The considered excitation is made of three ramps and is shown in the figure (right axis). The agreement is very good. It will be quantified throughout this study with the following standard quality metrics: RMS-error and fit level (in %) defined as

$$e_{RMS} = \sqrt{\frac{1}{n}L^2_{2\text{-}norm}(y - y_{ref})} \quad \text{and} \quad e_{fit} = 100\left(1 - \frac{L^2_{2\text{-}norm}(y-y_{ref})}{L^2_{2\text{-}norm}(y-\bar{y})}\right) \quad (5.5)$$

where $y$ and $y_{ref}$ will be defined along the text but refer always to model simulations outputs taken alternatively as the "measurement" signal ($y$) and the model response signal ($y_{ref}$). $\bar{y}$ refers to the mean of signal $y$, when averaged over the whole number of points (time steps) considered.

One can see in the insert that the step-LBTF approach is even more precise around discontinuities when compared to the full-LBTF. For this simulation, $\Delta t = 0.001s$, $N = 6283$ time steps, and we report in Table 2 the CPU times and errors calculated with the Step-LBTF simulation taken as the output reference $y_{ref}$. It is important to mention that the convolution approach (row 1) appears as the fastest (*) when considering the same time step as for the other methods. But this approach never supports the comparison in terms of precision. Generally, time steps more than 100 times smaller (line 1) are needed to obtain a solution with the same order of magnitude in precision. Therefore, in the next sections, results obtained with the convolution approach will be given for $\Delta t = 10^{-5}s$ to compare the different approaches for same accuracy levels. This constraint obviously precludes to use this approach for the purpose of discrete calculations on complex microstructures as it will never be efficient enough.

| Method | $\Delta t$ (s) | CPU time | $e_{RMS}$ (MPa) | $e_{fit}$ (%) |
|---|---|---|---|---|
| Convolution | $10^{-5}$ | 16.5 | 0.0835 | 100 |
|  | 0.001 | 0.014$^{(*)}$ | 0.154 | 100 |
| Full-LBTF |  | 0.064 | 0.396 | 99.999 |
| Step-LBTF |  | 0.101 | -- | -- |
| Step-ARX130 |  | 0.013 | 0.0062 | 100 |

**Table 2**: CPU times, RMS error and Fitting error for the different numerical approaches.

At this stage, the problem has been putted under incremental form with good accuracy, but requires the use of the past history values of the variables. This precludes the implementation in numerical codes where a space partition is required with a large amount of data to be stored: 12 components of the stress and strain tensors for each spatial node point and for each time step.

## 3 The step approach achieved with ARX parametric models.

ARX AutoRegressive models with eXogeneous inputs (Ljung, 1998) are parametric models which are shown in this paper to be very suited for being substituted to Laplace transforms and at the same time, providing an incremental approach of time domain computations. When the convolution problem requires the all input history, autoregressive models provide the model output at a given time as a linear combination of the only $na$ previous time steps of the output history and $nb$ time steps of the input history. Retardation or delay can be considered using $nk$ time steps. The structure of the model is given as

$$\sigma_{ARX}(t) = -\sum_{i=1}^{na} a_i \sigma_{ARX}(t - \tau_i) + \sum_{j=1}^{nb} b_j \varepsilon_{id}(t - \tau_{j+nk}) \qquad (5.6)$$

where $\tau_i = i\Delta t$ for $i \geq 1$, $\sigma_{ARX}(t) = 0$ for $t \leq 0$ and $\varepsilon_{id}(t)$ denotes the ideal (strain) excitation imposed on the (material REV) system. In command systems vocabulary, the stress figures the output and $\varepsilon_{id}$ the input or command signal. Coefficients $a_i, b_j$ are the observer Markov parameters. They have to be identified for a selected triplet $(na, nb, nk)$ – thus the common notation $ARX(na, nb, nk)$ used hereafter - during a so-called *calibration experiment*. Providing this latter to a specially designed optimization procedure (the functions made available by Ljung from the Identification Toolbox of Matlab will be used for this purpose (Ljung, 1988)), allows to obtain the optimal set of parameter vectors $a_i, b_j$ (dimensions and values) in the sense of a minimization of the residuals, while avoiding any over-adjustment. ARX model identifications are performed in 'simulation' mode, that is considering a cost function to minimize which is based on the residuals between the 'measured' and 'model' signals. The *calibration experiment* will be made here from a first simulation obtained in a direct way for a given strain excitation $\varepsilon^{\ell p}$ (selected from the test cases referenced in Table 2). Use can be made of either the step-LBTF or convolution approach to this aim, with a very small time-step if the latter is used. The second phase is a *validation experiment* which aims at quantifying the robustness of the identified ARX model i.e. its intrinsic character. This means ruling on the fact that parameters $a_i, b_j$ are now able, through the $ARX(na, nb, nk)$ model structure, to reproduce the real behavior in all situations. We will use the available test cases responses for various input excitations as data sets for the *validation experiment*.

It should be noted here that this ARX approach is deployed for a computing strategy but can be used by experimentalists. In that case a *calibration experiment* really obtained from a metrological set-up is necessary and the effect of the measurement noise has to be considered.

Practically, the identification of the *observer* Markov parameters will allow the computation of the *system* Markov parameters which correspond to a classical state-space representation of the system in control theory. This means that rather eq. (5.6), we will implement in the code the first-order matrix difference equation with $x(k)$ the state of the system, $\sigma(k)$ the output stress observable, $\varepsilon(k)$ the input strain.

$$\begin{aligned} x(k+1) &= Ax(k) + B\varepsilon(k) \\ \sigma(k) &= Cx(k) + D\varepsilon(k) \end{aligned} \tag{5.7}$$

Matrix $A$ is $n \times n$, $B$ is $n \times 1$ because only one input signal is considered, $C$ is $1 \times n$ because a single output signal is considered and $D$ is $1 \times 1$. These matrixes will be fully determined from identified observer Markov parameters and directly given with Matlab System Identification Toolbox. Formulation (5.7) is preferred in Matlab because formulation (5.6), although conceptually interesting, leads to a reconstruction of the validation experiment which depends from a deadbeat observer gain (not given by the Matlab System Identification Toolbox) and which is inherent to the compression mechanism that limits the input-output model (5.6) to a small number of parameters (Phan & Longman, 1996).

## 4 Simulation results on 3 VE behavior

In this section, we validate the step approach reached by substituting ARX models to the Laplace inversion of behavior's law formalized in Laplace domain. Three VE behaviors will be considered to assess the methodology which correspond to the three previously introduced behaviors: SLS, DLR and NIF models.

### 4.1 SLS Model

The response to the SLS model with parameters $E^u = 1000\ MPa, E^r = 100\ MPa, \tau_{SLS} = 1\ s$ is computed using Full-LBTF, convolution and Step-LBTF approaches. Figure 2 illustrates how well these approaches compare in test case 3 (3 successive ramps). We use then the Step-LBTF approach as the reference output $y_{ref}$ to identify an ARX model. The basic $ARX(1,3,0)$ is shown in Table 3 to produce very high level of precision. Of course, for a given calibration experiment (rows), the highest performances are obtained when the case considered for the validation experiment (columns) is the same (diagonal of Table 3). It is also clear that Case 4 used as calibration experiment (Calib4SLS) offers the best matching conditions for all other cases considered for the validation. Hence the identified ARX model used for the incremental strategy will be retained from this test case. For the validation test case Nr3 (Valid3SLS), figure 3 shows that the ARX model is perfectly able to compute the expected response incrementally using only three historical values of the input. Table 3 reports an excellent fitting performance of 99.98% with the smallest CPU times. Indeed, the $ARX(1,3,0)$ provides the same CPU times as the convolution method for the same time step when for accuracy, the RMS error resulting from the convolution product approach is 25-times greater than the ARX approach. The same applies for all other validation cases.

| Validation → ↓Calibration | Valid0SLS | Valid1SLS | Valid2SLS | Valid3SLS | Valid4SLS |
|---|---|---|---|---|---|
| Calib0SLS | 99.68 | 96.89 | 96.5 | 98.81 | 95.14 |
| Calib1SLS | 99.06 | 99.97 | 99.85 | 99.93 | 99.84 |
| Calib2SLS | 98.95 | 98.26 | 99.99 | 99.89 | 99.84 |
| Calib3SLS | 98.9 | 95.91 | 99.72 | 99.98 | 99.63 |
| Calib4SLS | 99 | 96.49 | 99.83 | 99.98 | 99.98 |

**Table 3:** Fitting error obtained for the different test cases (0 to 4) and for the different set of **Calib**ration/**Valid**ation numerical experiments for the SLS behavior's law

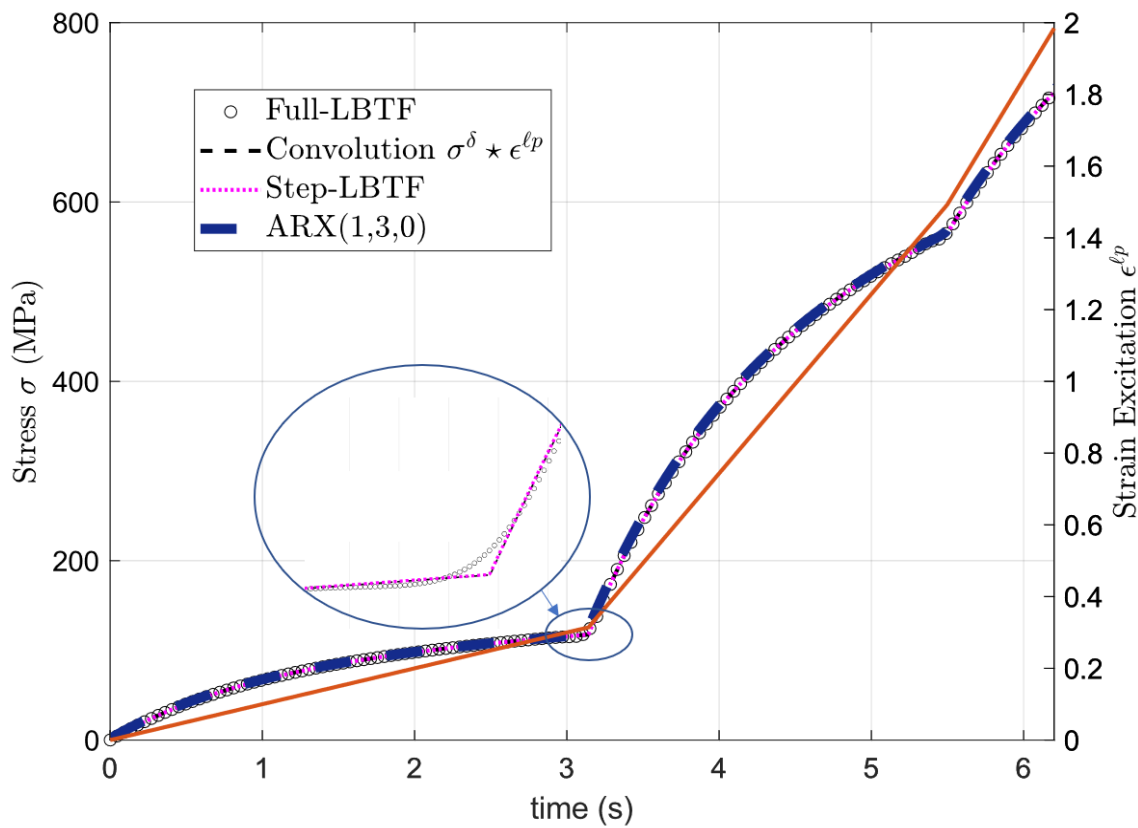

**Figure 2:** ARX and LBTF approaches compared in the SLS behavior case, with excitation made of 3 ramps (Case3). Calibration of $ARX(1,3,0)$ model based on Case4.

## 4.2 DLR Model

We consider now a more complex viscoelastic behavior based on the DLR model of Equation (5.1) with a relaxation spectrum described over $d = 6$ decades, with a discrete set of N=50 relaxation modes and parameters defined like for the SLS simulation: $E^u = 1000\ MPa, E^r = 100, \tau_{max} = 1s$. The Step-LBTF approach is chosen again to identify the two $ARX(1,3,0)$ and $ARX(2,4,0)$ models considered here. Table 4 gives the fitting error obtained after identification of $ARX(2,4,0)$ model. It is still clear that Test Case 4 offers a signal with sufficiently broad information in frequency domain to cover correctly other types of excitation. Table 4 also illustrates that for this VE behavior, fitting performances, if still excellent, decrease when compared to the simple SLS behavior. That is also the reason why the simple $ARX(1,3,0)$ structure now fails in describing accurately the response in various other situations. Figure 3 compares all approaches in test case 1: the input excitation is a crenel of duration $t_c = 3.2s$ and amplitude $\varepsilon_0 = 1$ (figure 3 - Right axis). Figure 4 plots the residuals obtained for the different methods (always considering the Step-LBTF approach for $y_{ref}$).

| Validation → <br> ↓Calibration | Valid0DLR | Valid1DLR | Valid2DLR | Valid3DLR | Valid4DLR |
|---|---|---|---|---|---|
| Calib0DLR | 99.36 | 89.56 | 96.74 | 96.89 | 91.09 |
| Calib1DLR | 97.7 | 96.8 | 99.15 | 99.15 | 96.13 |
| Calib2DLR | 98.05 | 87 | 99.93 | 99.93 | 99.64 |
| Calib3DLR | 97.99 | 87.43 | 99.92 | 99.93 | 99.54 |
| Calib4DLR | 98.14 | 86.39 | 99.85 | 99.85 | 99.82 |

**Table 4:** $e_{fit}$ errors (in %) obtained when using the different test cases for calibration and validation experiments to identify the $ARX(2,4,0)$ model.

| Method : | | Full_LBTF | Convolution | Step_LBTF | ARX130 | ARX240 |
|---|---|---|---|---|---|---|
| $\varepsilon^{\ell p}$ − case0 | CPU | 0.08 | 21.8 | 0.19 | 0.006 | 0.011 |
| | $e_{fit}$(%) | 99.96 | 98.96 | - | 99.88 | 99.96 |
| | $e_{RMS}$(MPa) | 1.53 | 1.41 | - | 2.6 | 1.52 |
| $\varepsilon^{\ell p}$ − case1 | CPU | 0.074 | 1.062 | 0.161 | 0.005 | 0.008 |
| | $e_{fit}$(%) | 98.68 | 99.94 | - | -112 | 95.24 |
| | $e_{RMS}$(MPa) | 22.98 | 4.6 | - | 291 | 43.7 |

**Table 5:** Comparison of LBTF and ARX approaches for different excitations (time step $\Delta t = 10^{-3}s$ except for the convolution $\Delta t = 10^{-5}s$).

The sudden input step induces an instantaneous response. A relaxation stage is then observed and fully completed here because $t_c > \tau_{max}$ before it restarts oppositely at $t_c$ when the applied excitation ceases. It can be observed, especially at discontinuities, that for this more complex behavior's law involving a spectrum of relaxation times, the $ARX(2,4,0)$ has superior abilities than $ARX(1,3,0)$. This is clearer from the values of errors reported in Table 4 and the graph of residuals in Figure 4. Because the ARX mathematical structure does not change and involves very few time steps, the CPU times are of same order (Table 5), much less than other approaches, especially the Step-LBTF which requires the full history of the input command.

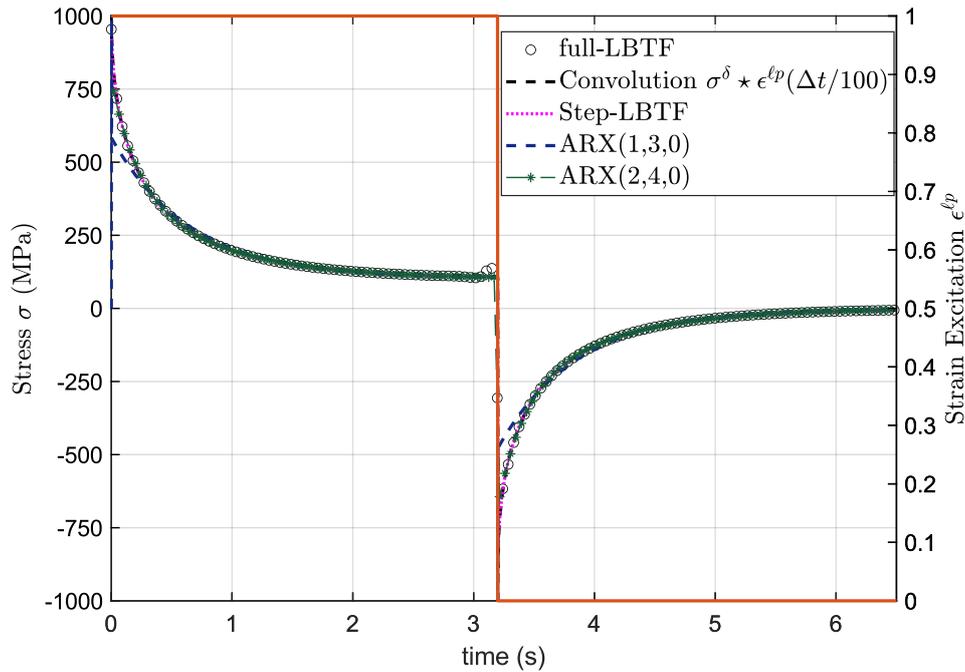

**Figure 3:** Case 1 - Crenel strain excitation with DLR viscoelastic behavior (Valid1DLR). Comparison of LBTF approaches and ARX 130 and ARX240 parametric models identified on test case 4 (Calib4DLR).

For the smooth excitation (case 0), figure 5 reports the stress output for all approaches and Table 5 reports the achieved performances. In that case, both $ARX$ identified models give similar performances. The RMS errors of 2.6 and 1.5 MPa for respectively the $ARX(1,3,0)$ and $ARX(2,4,0)$, when compared to the maximum signal amplitude of 250MPa correspond to about 1% and 0.6%. They were 16.5 and 2.5% for the crenel excitation. $ARX(2,4,0)$ can then be favored in case of expected discontinuities in the strain input path, for nearly equal and small CPU times in any case. The residuals between the reference Step-LBTF signal and all other approaches (not shown here) behave similarly as those shown in Figure 4 for the crenel excitation, except where discontinuities take place.

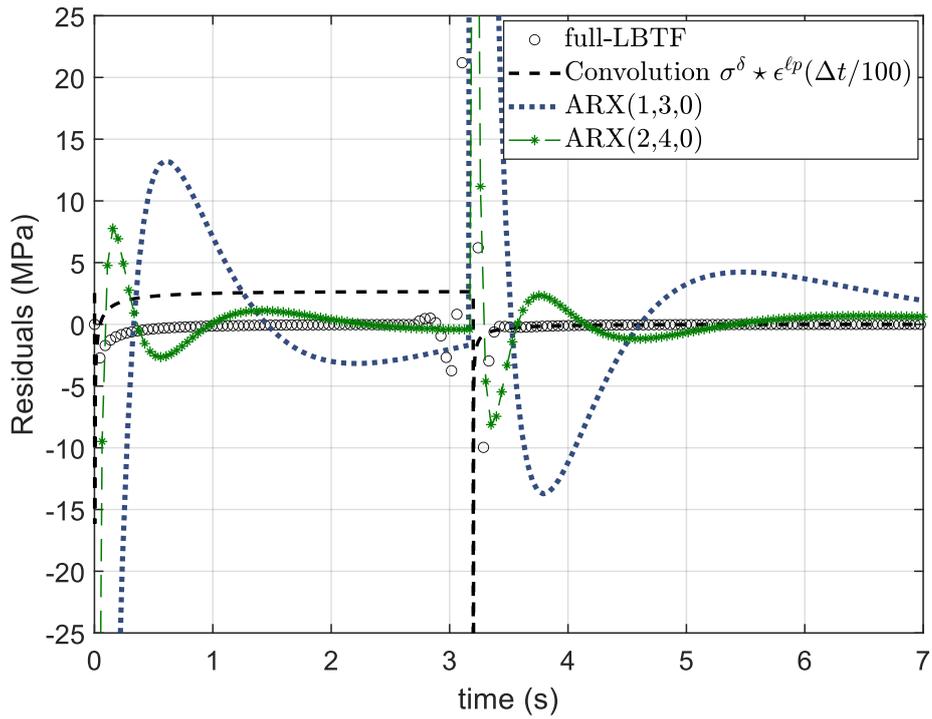

**Figure 4:** Residuals corresponding to Figure3. The Step_LBTF "signal" is used as reference for the calibration of ARX models.

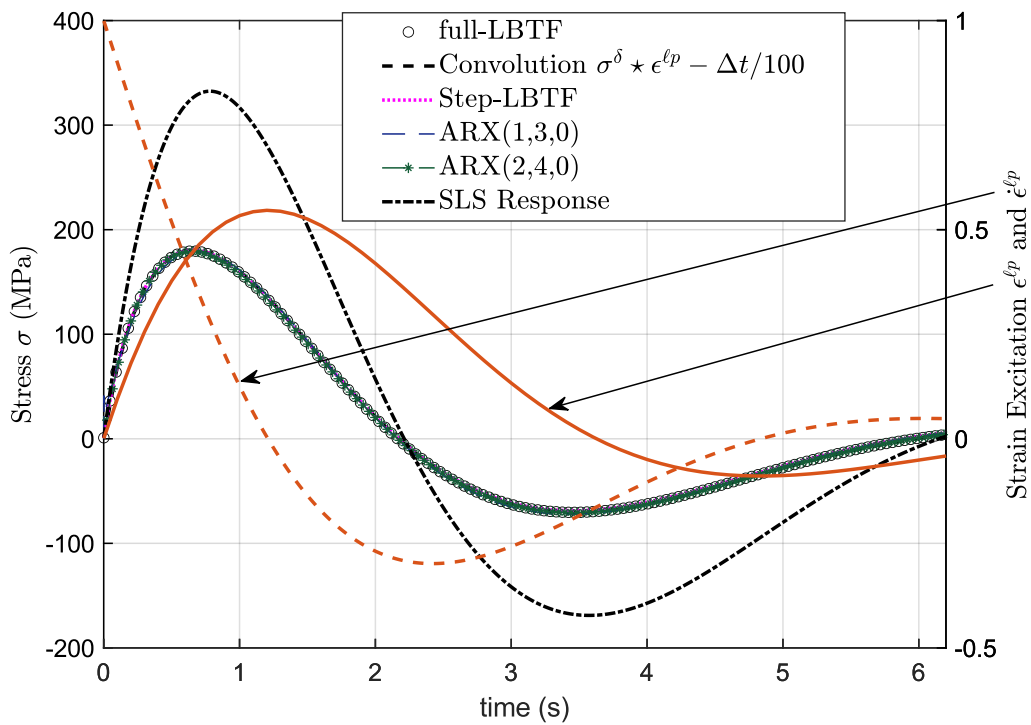

**Figure 5:** ARX and LBTF approaches compared in the DLR behavior case, with excitation made of the smooth excitation $\varepsilon^{\ell p}$=f(t) (Case0). Calibration of *ARX* models on Case4.

## 4.3 NIF Models

Finally, the so-called fractional models are considered i.e. models involving a frequency LBTF with non-integer exponents in terms of the Laplace (or frequency) variable. These later imply a non-integer differentiation operator in temporal domain. We will consider two of them, having distinct conceptual basis.

### 4.3.1 Non-integer Oustaloup model

The first considered model was proposed initially by A. Oustaloup (Sabatier et al., 2015) to represent recursive schemes of the impulse response of a Linear Time Invariant system whose mathematical structure is made of Prony series with recursive factors. We proved that such a model can be related to the DLR approach in the asymptotic limit of infinite modes of relaxation (André et al., 2003). The formal expression of the behavior's law is known directly in Laplace domain (Eq.5.8)

$$\bar{Y}_n(p) = \frac{\bar{\sigma}(p)}{\bar{\varepsilon}(p)} = E^u + (E^r - E^u)\frac{\left(1 + \frac{p}{\omega_{high}}\right)^{n-1}}{\left(1 + \frac{p}{\omega_{low}}\right)^n} \qquad (5.8)$$

with $\omega_{high} = 1/\tau_{min}$, $\omega_{low} = 1/\tau_{max}$ defining the bounds of a spectrum of relaxation times covered by the Laplace variable range.

The values of $E^u = 1000$, $E^r = 100$ in $MPa$ are the same as those taken for the previous SLS and DLR behaviors. The relaxation component of the $\bar{Y}_n$ transfer function depends on the last three parameters appearing in Equation (5.8). We will consider the case of $n = 0.62$, $\tau_{min} = 1\mathrm{e}-2\ \mathrm{s}$, $\tau_{max} = 16\ \mathrm{s}$.

This transfer function is rather elaborated (see Andre 2003 for details). In the $\omega_{high} \rightarrow \infty$ limit, it reduces to the well-known Davidson-Cole approach in frequency domain. In temporal domain, this transfer function implies non-integer derivatives. Therefore, the simulation presented in this section with such a transfer function and its "reduction" to ARX models will illustrate how non-integer models of viscoelasticity, formulated in Laplace domain, can be used easily in incremental time domain computations.

Tables 6 and 7 give the $e_{fit}$ errors obtained for the different possible combinations of the set of calibration and validation simulations for both $ARX(1,3,0)$ and $ARX(2,4,0)$ respectively. The $ARX(2,4,0)$ model appears of course superior to the $ARX(1,3,0)$. Again, one must point out when comparing the results of line 2 versus column 2 in both tables that it is preferable to consider the test case 1 (the crenel) as calibration test to qualify the ARX model. Its performance is then better for the other validation cases than the contrary.

Table 8 gathers CPU times, $e_{fit}$ and RMS errors for different test cases with the specified ARX calibration model. The $e_{fit}$ and RMS errors are established with reference taken from the Step_LBTF. The residuals RMS error given in the 6[th] and last sub-line corresponds to test Case 1 and 4 respectively and can be compared by looking at the signal response variation shown in Figures 6 and 7. CPU times for ARX models are greatly diminished, by an order of magnitude of 10 in average between the Full_LBTF and $ARX(2,4,0)$ model. Looking at lines 3 and 4 (same test case 4 but calibration of ARX models from case 1 or 4) show again that the calibration performed with the reference crenel excitation gives results as good as for the proper test case.

| Validation → ↓Calibration | Valid0Yn | Valid1Yn | Valid3Yn | Valid4Yn |
|---|---|---|---|---|
| Calib0Yn | 98.11 | 36.77 | 99.31 | 97.08 |
| Calib1Yn | 96.36 | 95.13 | 99.67 | 98.32 |
| Calib3Yn | 95.44 | 61.2 | 99.87 | 95.01 |
| Calib4Yn | 97.67 | 37.12 | 99.7 | 99.31 |

**Table 6:** $e_{fit}$ errors (in %) obtained when using the different test cases for calibration and validation experiments to identify the $ARX(1,3,0)$ model.

| Validation → ↓Calibration | Valid0Yn | Valid1Yn | Valid3Yn | Valid4Yn |
|---|---|---|---|---|
| Calib0Yn | 99.78 | 91.37 | 99.39 | 97.52 |
| Calib1Yn | 98.53 | 98.5 | 99.38 | 99.16 |
| Calib3Yn | 99.02 | 90.93 | 99.99 | 99 |
| Calib4Yn | 99.28 | 88.54 | 99.96 | 99.93 |

**Table 7:** $e_{fit}$ errors (in %) obtained when using the different test cases for calibration and validation experiments to identify the $ARX(2,4,0)$ model.

| Method : | | Full_LBTF | Step_LBTF | ARX130 | ARX240 |
|---|---|---|---|---|---|
| $\varepsilon^{\ell p}$ − case0 Calib0 | CPU | 0.2034 | 0.1325 | 0.0054 | 0.01 |
| | $e_{fit}(\%)$ | 99.99 | - | 99.64 | 99.9995 |
| | $e_{RMS}$(MPa) | 1.7 | - | 3.37 | 0.4 |
| $\varepsilon^{\ell p}$ − case1 Calib1 | CPU | 0.085 | 0.126 | 0.005 | 0.0076 |
| | $e_{fit}(\%)$ | 99.6 | - | 99.65 | 99.87 |
| | $e_{RMS}$(MPa) | 26.4 | - | 23.9 | 14.7 |
| $\varepsilon^{\ell p}$ − case4 Calib1 | CPU | 0.08 | 0.25 | 0.006 | 0.009 |
| | $e_{fit}(\%)$ | 99.999 | -- | 99.97 | 99.99 |
| | $e_{RMS}$(MPa) | 0.54 | -- | 5.8 | 3.3 |
| $\varepsilon^{\ell p}$ − case4 Calib4 | CPU | 0.079 | 0.23 | 0.0065 | 0.009 |
| | $e_{fit}(\%)$ | 99.999 | -- | 99.99 | 100 |
| | $e_{RMS}$(MPa) | 0.54 | -- | 2.36 | 0.25 |

**Table 8:** Comparison of LBTF and ARX approaches for different excitations (time step $\Delta t = 10^{-3}s$) and ARX calibration models.

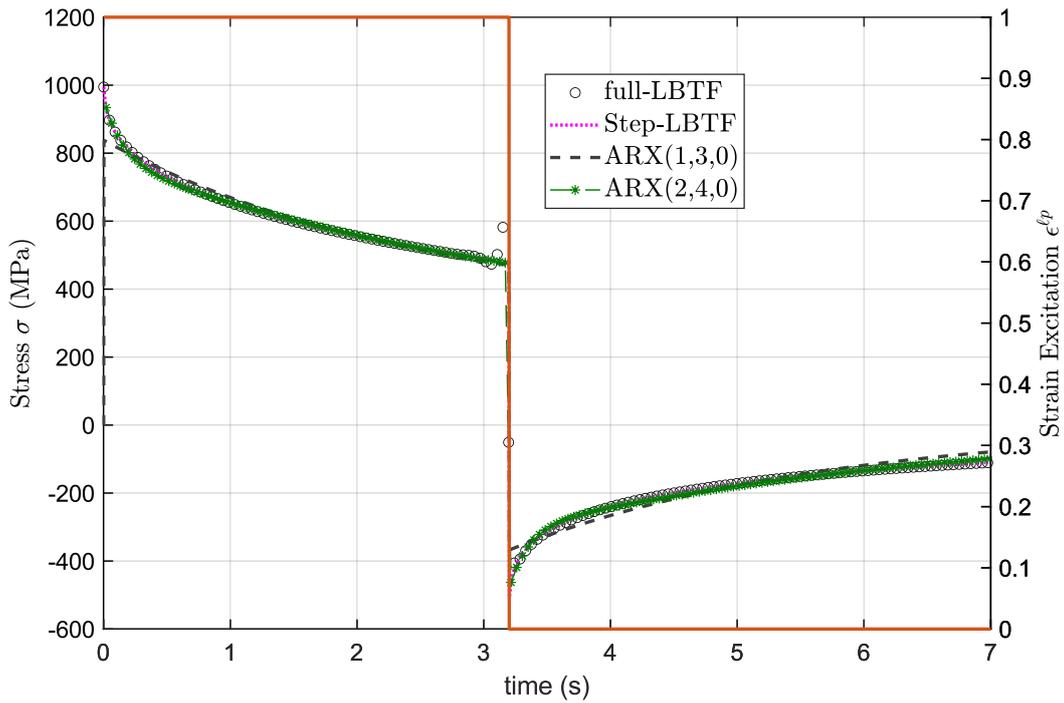

**Figure 6:** ARX and LBTF approaches compared in the fractional behavior case, with the crenel excitation $\varepsilon_1^{\ell p}$ (Case1). Calibration of *ARX* models on Case1.

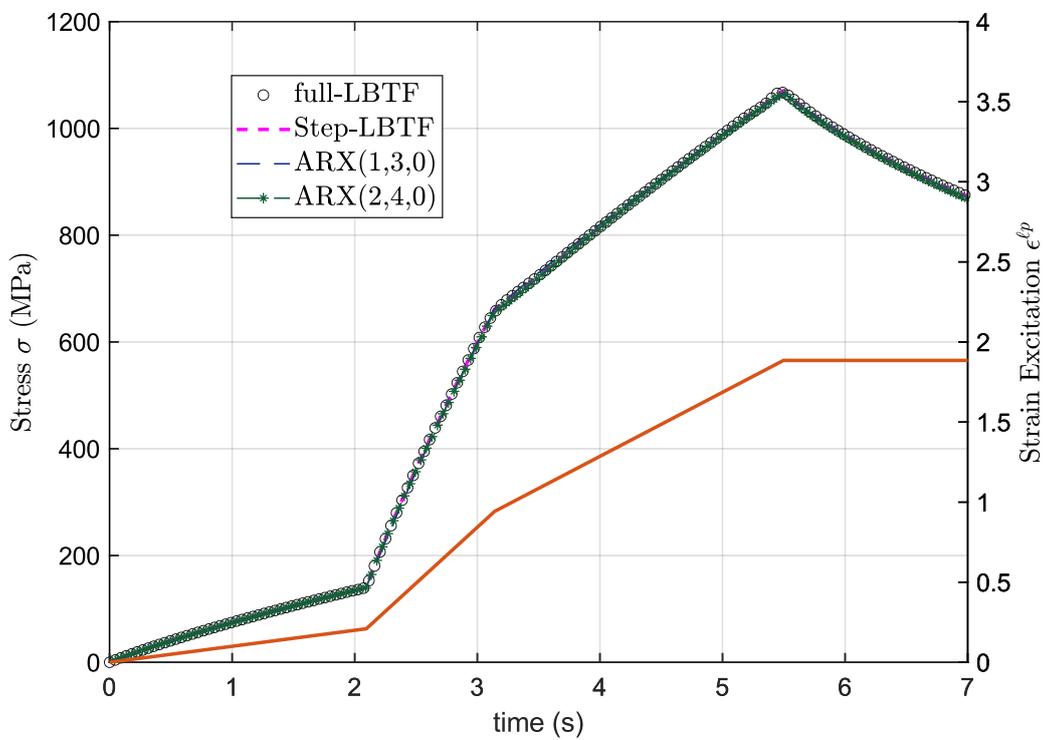

**Figure 7:** ARX and LBTF approaches compared for the NIF behavior case, with excitation made of the 3 ramps+dwell excitation $\varepsilon_4^{\ell p}$ (Case4). Calibration of *ARX* models on Case4.

### 4.3.2 Fractional Rabotnov model

The second considered model relies on the use of a relaxation kernel noted $\beth_\alpha^* (\beta, t)$ which was proposed by Rabotnov and Scott-Blair (Sevostianov et al., 2015). Like the previous one, it is based on a very tractable analytical expression in Laplace domain which leads to the following LBTF:

$$\bar{Y}_\alpha(p) = \frac{\bar{\sigma}(p)}{\bar{\varepsilon}(p)} = E^u + (E^r - E^u)\mathcal{L}[\beth_\alpha^* (\beta, t)]$$
$$= E^u + (E^r - E^u)\frac{1}{p^{1-\alpha} + \beta} \quad (5.9)$$

It does not correspond to the same mathematical structure as given by Equation (5.8). In the above expression, $\alpha$ denotes the non-integer exponent, $\beta$ the inverse of a (relaxation) time, and the Rabotnov kernel is defined in temporal domain by

$$\beth_\alpha^* (\beta, t) = E_{\alpha+1,\alpha+1}(\beta t^{\alpha+1}) = t^\alpha \sum_{n=0}^{\infty} \frac{(\beta t^{1+\alpha})^n}{\Gamma[(n+1)(1+\alpha)]} \quad (5.10)$$

This kernel is expressed through the 2-parameters Mittag-Leffler function. Figure 8 below gives an example of the Mittag-Leffler function calculated from Equation (5.10) and by performing De Hoog's inverse Laplace algorithm on the second term of the rhs of Equation (5.9). Mittag-Leffler are evident functions arising from irreversible dynamics like relaxation phenomena in disordered systems under various physical approaches of 'slow-dynamics' type. For example in (Weron & Kotulski, 1996) the Fractal Time Random Walk model of dynamical phenomena in complex condensed matter systems is shown to produce the Mittag-Leffler function, as the inverse Fourier transform of the Cole-Cole function. When random walks are used to approach anomalous diffusion or relaxation in complex systems, the introduction of statistically distributed waiting times before each walk step has been also shown to produce dynamics described by ML functions (Metzler & Klafter, 2000). They appear as the formal solution of Fokker-Planck equations used in a subdiffusion (i.e. non Brownian) phenomenon (Metzler & Klafter, 2002).

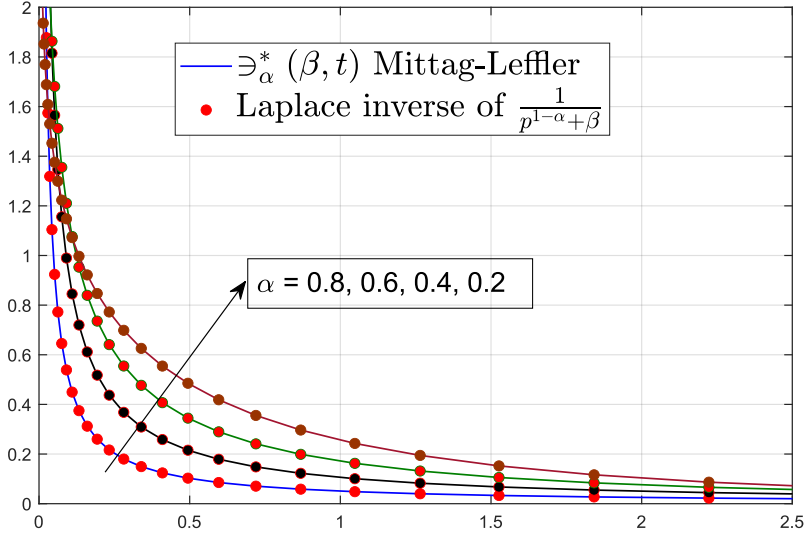

**Figure 8:** Rabotnov Kernel – Comparison of the time expression $\beth_\alpha^*(\beta,t)$ defined as a Mittag-Leffler function (5.10) and the Laplace original of $\frac{1}{p^{1-\alpha}+\beta}$ ($\alpha = 0.8$ to $0.2$, $\beta = 1$ i.e. characteristic relaxation time of $1s$).

Concluding the study with this example presents the interest of being able to produce the temporal expression of the impulse response and then a convolution product just as it was possible for the SLS and DLR models but not with the NIF 'Oustaloup' model. We simply have in that case for the 'Rabotnov' impulse response:

$$\sigma^\delta(t) = \mathcal{L}^{-1}\bar{H}(p)(t) = E^u \delta(t) + (E^r - E^u) t^{-\alpha} E_{1-\alpha, 1-\alpha}\left(-\frac{1}{\tau} t^{1-\alpha}\right) \qquad (5.11)$$

As a recall, this leads for any loading path $\varepsilon^{\ell p}(t)$ to compute the convolution product $\sigma(t) = (\sigma^\delta * \varepsilon^{\ell p})(t)$. Within a full identical strategy, we just show in Figure 9 the performance of $ARX(1,3,0)$ model in this new case of wide interest in the field of relaxation phenomena. $ARX(1,3,0)$ parameters were identified from the simulated outputs to $\varepsilon_4^{\ell p}$ (Case4 of the multisequence loading path). Simulations were performed for the same values of $E^u = 1000\ MPa$, $E^r = 100\ MPa$ as before, with $\beta = 1\ s^{-1}$ and the four $\alpha$ values $0.2, 0.4, 0.6, 0.8$ considered in Figure 8. The identified $ARX(1,3,0)$ models gave respectively a fit error of $98.05\%, 98.3\%, 99\%$ and $99.58\%$ and were then used to compute the output to $\varepsilon_3^{\ell p}$. Figure 9 reports the 4 sets of curves obtained to compare the Step-LBTF and the ARX responses. Even with this low parameterized ARX structure, all approaches (LBTF, Convolution, Step-LBTF, ARX) compare again very well in all cases (error fit above $99.99\%$).

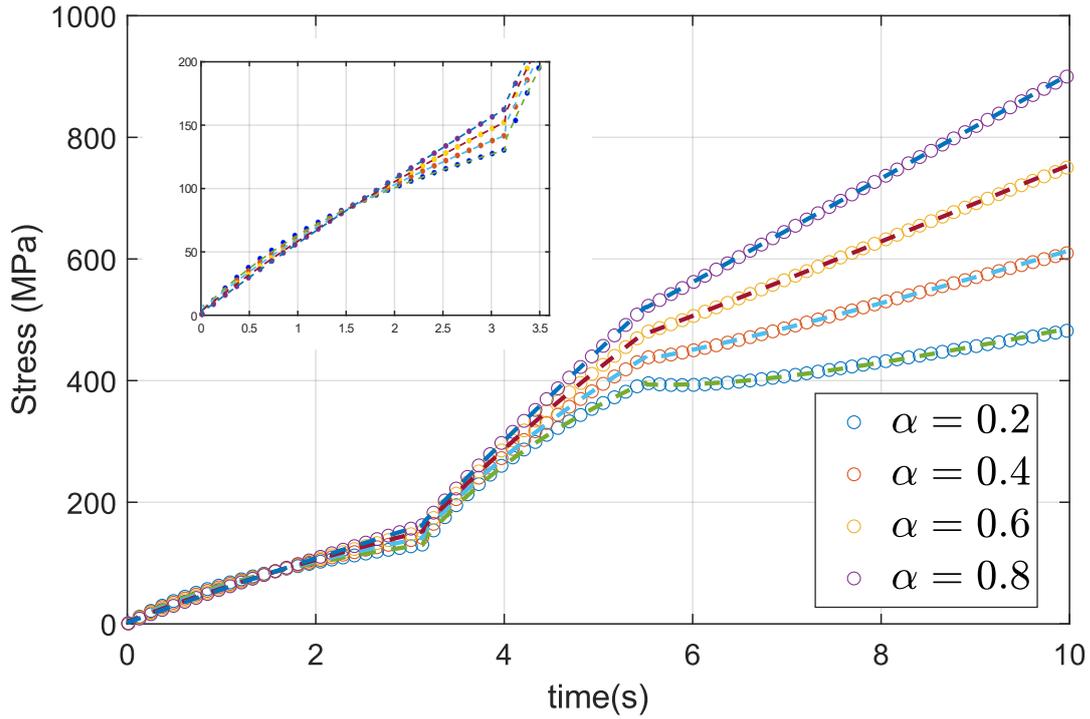

Figure 9: $ARX(1,3,0)$ (doted lines) and Step-LBTF approaches (circle dots) compared for the fractional 'Rabotnov' VE kernel, with excitation $\varepsilon_3^{\ell p}$, made of 3 ramps (Case3). $ARX130$ models calibrated on Case4.

## 5 Conclusion and perspectives

We presented a strategy for solving the quasi-static mechanical problem under any kind of excitation for a viscoelastic material which constitutive law is prescribed in Laplace domain but immediately substituted by parametric models. For the simple problem of a homogeneous body, ARX models could provide an interesting alternative to the use of the exact models in the case where very short computation times are required. For example, an UMAT subroutine in Abaqus could work as a generic behavior's law with the ARX parameters being introduced by the user according to the ARX structure he decided to use. ARX models have been shown here to perform remarkably well for all types of responses in a mechanical VE problem, in the linearity domain w.r.t. the input signals. But the more promising interest lies probably in the use of ARX parametric models as proxy models for behavior's laws defined in terms of their Laplace transform (in the frequency domain as produced by DMA for example) that cannot be readily used in incremental schemes of numerical codes. Therefore, it could provide straightforward alternatives in the case of heterogenous media with viscoelastic phases to produce the homogenized response of the REV or to simulate a complex structure.

# Appendix

**Numerical strategy based on the Duhamel theorem and URSS-response.**

We consider the problem $\sigma(t) = \mathcal{L}^{-1}\{\bar{H}(p)\varepsilon^{\ell p}(p)\}$ with a solution given in incremental manner $\sigma(t + \Delta t) = \sigma(t) + \Delta\sigma^t$ where $\Delta\sigma^t$ will be computed from the knowledge of the discretized URSS-Response $\sigma_0^{\nearrow-}(t) = \mathcal{L}^{-1}\{\bar{H}(p)\frac{1}{p^2}(1 - e^{-p\Delta t})\}$ from one side and the piecewise discretization of $\varepsilon^{\ell p}(t)$ in terms of staggered ramp functions from the other side.

Example can be taken with a signal composed of two ramps as shown in figure 1.

$\Delta\sigma^t$ will be the sum of the contribution of the URSS-Response at initial time and of the successive contributions due to the step-changes in strain rates, worked out through the scalar product realizing the convolution. After defining:

- a vector of time $\boldsymbol{td} = \{t_1, t_2 = t_1 + \Delta t, t_3 = t_1 + 2\Delta t, \cdots, t_N = t_1 + (N-1)\Delta t\}$,
- a vector of the *strain rate* jumps $\boldsymbol{Sjump} = \{\Delta S_1, \Delta S_2, \Delta S_3, \cdots, \Delta S_n\}$ calculated from the piecewise linearly discretized strain loading path (see Figure 1) producing the vector of slopes $\boldsymbol{Slope} = \{S_1, S_2, \cdots, S_k = (\varepsilon_{k+1}^{\ell p} - \varepsilon_k^{\ell p})/\Delta t, \cdots, S_n\}$ and
- a vector of the discretized URSS-Response initiated at a 0-initial time (equation 5.4) $\boldsymbol{Runit} = \{\sigma_0^{\nearrow-}(t_1), \sigma_0^{\nearrow-}(t_2), \sigma_0^{\nearrow-}(t_3), \cdots, \sigma_0^{\nearrow-}(t_n)\}$,

the application of Duhamel's theorem readily brings this simple algorithm:

```
Stress_init = Slope(1) * Runit
# Stress response to initial strain jump for all time steps

Stress(1) = Stress_init(1)
# Initialization  σ(t₁) = S₁ × σ₁↗⁻   S₁: Slope of excitation on
first time step.

Loop over k = 1: N-1

    A1 = dot(Sjump (1:k-1), Runit (k:-1:2))    #Scalar products
    A2 = dot(Sjump (2:k), Runit (k-1:-1:1))

    Stress(k + 1) = Stress(k) + Stress_init(k) + (A1 + A2)/2
    #Midpoint to enhance precision

End Loop
```